\documentclass[12pt]{article}

\usepackage[utf8]{inputenc}
\usepackage{slashed}
\usepackage{jheppub}
\pdfoutput=1

\usepackage{color}
\usepackage{graphicx}   
\usepackage{bm}
\usepackage{amsmath}
\usepackage{amsfonts}
\usepackage{eufrak}
\usepackage{hyperref}

\def\sumint{\,\hbox{$\sum$}\!\!\!\!\!\!\!\int}

\begin{document}

\title{Strong Coupling Universality at Large N for Pure CFT  Thermodynamics in 2+1 dimensions}

\author{Oliver DeWolfe$^{1,2}$ and}
\emailAdd{oliver.dewolfe@colorado.edu}
\author{Paul Romatschke$^{1,2}$}
\affiliation{$^1$ Department of Physics, University of Colorado, Boulder, Colorado 80309, USA}
\affiliation{$^2$ Center for Theory of Quantum Matter, University of Colorado, Boulder, Colorado 80309, USA}
\emailAdd{paul.romatschke@colorado.edu}

\abstract{
  Pure CFTs have vanishing $\beta$-function at any value of the coupling. One example of a pure CFT is the O(N) Wess-Zumino model in 2+1 dimensions in the large N limit. This model can be analytically solved at finite temperature for any value of the coupling, and we find that its entropy density at strong coupling is exactly equal to $\frac{31}{35}$ of the non-interacting Stefan-Boltzmann result. We show that a large class of theories with equal numbers of N-component fermions and bosons, supersymmetric or not, for a large class of interactions, exhibit the same universal ratio. For unequal numbers of fermions and bosons we find that the strong-weak thermodynamic ratio is bounded to lie in between $\frac{4}{5}$ and $1$.
  }

\maketitle

\section{Introduction}

Among quantum field theories, conformal field theories (CFTs) are special because their $\beta$-function vanishes and no mass scales are present. Typically, CFTs can be obtained from a quantum field theory parent by tuning the coupling to a critical value. However, there is a class of special CFTs, sometimes referred to as ``pure'' CFTs, which are CFTs for \textit{all} values of the coupling, not just at  a finite number of critical points. Pure CFTs are not common, because they are special examples of an already special set of quantum field theories. One example is ${\cal N}=4$ Super-Yang-Mills theory in 3+1 dimensions, which is a pure CFT for any number of colors $N$. However, one may also have theories that are pure CFTs only in the limit $N \to \infty$, such as the interacting O(N) model in 2+1 dimensions.

Because pure CFTs are so special, it is interesting to study some of their properties. A famous property of ${\cal N}=4$ SYM in 3+1 dimensions is that its free energy at finite temperature evaluated at infinite coupling (calculated via its conjectured gravity dual \cite{Maldacena:1997re}) is exactly $3/4$ of the Stefan-Boltzmann value expected for a free theory \cite{Gubser:1998nz}. Using the strong constraints of CFTs in 1+1 dimensions \cite{Belavin:1984vu}, it is possible to construct pure CFTs with a strong-weak thermodynamic ratio given by the ratio of central charges, finding a strong-weak thermodynamic ratio of $\frac{1}{2}$ for scalar $\phi^4$ theory in 1+1 dimensions with a mass tuned to the Ising point \cite{inprep}. Finally, for scalar theory with quartic interaction in 2+1 dimensions, a strong-weak thermodynamic ratio of 4/5 has been found in Ref.~\cite{Drummond:1997cw} (cf. Ref.~\cite{Sachdev:1993pr}), and the ratio has recently been shown to be universal in a large class of pure CFTs with only bosonic degrees of freedom\cite{Romatschke:2019ybu}.

How general is the ratio of 4/5 for pure CFTs in 2+1 dimensions? In order to answer this question, we consider generalizing Ref.~\cite{Romatschke:2019ybu} to include fermionic degrees of freedom in this work. We will start by considering a supersymmetric theory in 2+1 dimensions, the O(N) Wess-Zumino model, but then start modifying the Lagrangian, first by breaking supersymmetry, then by changing the interactions, and finally by considering unequal numbers of fermionic and bosonic degrees of freedom.

\section{Calculation}

Let us consider the O(N) supersymmetric Wess-Zumino model \cite{Wess:1974tw} in 2+1 dimensions given by the superspace action
\begin{equation}
  S=\int d^3x d\bar\theta d\theta \left(\frac{1}{2}\bar D\Phi_a D\Phi_a + \frac{2\lambda}{N} \left(\Phi_a \Phi_a\right)^2\right)\,,
\end{equation}
where the superfield $\Phi_a$ and its spinorial differentiation can be written in components as
\begin{eqnarray}
  \Phi_a&=&\phi_a+\bar\theta^\alpha \psi_{\alpha a}+\frac{1}{2}\bar\theta \theta F\,,\\
  D_\alpha \Phi_a &=&\psi_{\alpha a}+\theta_\alpha F_a-i \left(\gamma^\mu \theta\right)_\alpha \partial_\mu \phi_a+\frac{i}{2}\bar\theta \theta \slashed{\partial}\psi_{a \alpha}\,.
\end{eqnarray}
Here $\phi_a,F_a$ are N-component real scalar fields and $\psi_a$ is an N-component Majorana spinor in 3 dimensions. Performing the integration over supervariables $\bar \theta, \theta$ and integrating out the scalars $F_a$ in the action leads to
\begin{equation}
  S=\int d^3x \left(\frac{1}{2}\partial_\mu \phi_a \partial^\mu \phi_a
  +\frac{1}{2}\bar\psi_a i\slashed{\partial}\psi_a
  -\frac{8\lambda^2}{ N^2} \left(\phi_a\phi_a\right)^3 -\frac{\lambda}{N}\bar\psi_a \psi_b \left(4 \phi_a\phi_b+2 \delta_{ab}\phi_c\phi_c\right)\right)\,.
  \end{equation}
The Yukawa-type terms break the parity symmetry $x^2 \to - x^2$, $\psi^a \to i \gamma^2 \psi^a$, but this indicates the physics of the model is invariant under $\lambda \to -\lambda$ up to a parity transformation. The term $\bar\psi_a \psi_b \phi_a\phi_b$ does not contribute to leading order in the large N limit and hence will be neglected in the following.

The theory can be heated up to temperature $T$ by analytically continuing time to be imaginary and compactified on a circle of radius $\beta=T^{-1}$. The partition function is then given by
\begin{equation}
  Z=\int {\cal D}\phi_a {\cal D}\psi_a e^{-S_E}\,,
  \end{equation}
with $S_E$ the Euclidean action
\begin{equation}
  \label{eq:seuclid}
  S_E=\int d^3x \left(\frac{1}{2}\partial_\mu \phi_a \partial_\mu \phi_a
  +\frac{1}{2}\bar\psi_a \slashed{\partial}\psi_a
  +\frac{8\lambda^2}{ N^2} \left(\phi_a\phi_a\right)^3 + \frac{2\lambda}{N} \bar\psi_a \psi_a \phi_b\phi_b\right)\,.
  \end{equation}

Introducing an auxiliary field $\sigma=\phi_c\phi_c/N$ and its Lagrange multiplier $\zeta$ as {$1=\int {\cal D}\sigma {\cal D}\zeta e^{i\int \zeta(\sigma-\phi_c\phi_c/N)}$, only the zero modes of $\sigma,\zeta$ contribute to the leading order large N result of the partition function. As a result, the Gaussian integral over the bosonic and fermionic fields may be performed as
\begin{eqnarray}
  \int {\cal D}\phi_a e^{-\int d^3x \phi_a \left(-\partial^2+m^2\right) \phi_a}&=&e^{-\frac{\beta V N}{2} \sum\!\!\!\!\! \int_P \ln (P^2+m^2)}\equiv e^{-\beta V N J_B(m)}\,.\\
  \int {\cal D}\psi_a e^{-\int d^3x \psi_a \left(\slashed{\partial}+m\right) \psi_a}&=&e^{\frac{\beta V N}{2}\sum\!\!\!\!\! \int_{\{P\}} \ln (P^2+m^2)}\equiv e^{\beta V N J_F(m)}\,,
\end{eqnarray}
where bosonic and fermionic sum-integrals in $3-2\epsilon$ dimensions are written as $\sumint_{\  K}=T \sum_n \int \frac{d^{2-2\epsilon}{\bf k}}{(2 \pi)^{2-2\epsilon}}$ with $K,\{K\}$ denoting summation over bosonic (fermionic) Matsubara modes $\omega_n=2 \pi n T$, $\{\omega_n\}=2 \pi T(n+\frac{1}{2})$ (see e.g. Ref.~\cite{Laine:2016hma}).

Rescaling $\zeta\rightarrow N \zeta/2$ and $\sigma\rightarrow \sigma/(4 \lambda)$ this leads to the partition function given by
\begin{equation}
  Z=\int d\sigma d\zeta e^{-\beta V N\left(\frac{\sigma^3}{8\lambda}-\frac{i\zeta\sigma}{8\lambda} +J_B(\sqrt{ i \zeta})-J_F(\sigma)\right)}\,,
\end{equation}
where the bosonic and fermionic thermal sum-integrals $J_B,J_F$ in 2+1 dimensions are finite in the $\epsilon\rightarrow 0$ limit.
For large N, the partition function may be evaluated exactly using the saddle points located at $i\zeta=z^*,\sigma=\sigma^*$ given by the solution of the non-perturbative coupled gap equations 
\begin{equation}
  \label{eq:saddles}
  \frac{z^*}{4\lambda}=\frac{3 \sigma^2}{4\lambda}-2\sigma I_F(\sigma)\,,
  \quad
  \frac{\sigma^*}{4\lambda}=I_B(\sqrt{z^*})\,,
\end{equation}
where $I(m)\equiv 2\frac{d J(m)}{dm^2}$ with
\begin{equation}
  \label{eq:is}
  I_F(m)=-\frac{m}{4\pi}-\frac{T}{2\pi}\ln\left(1+e^{-m/T}\right)\,,\quad
  I_B(m)=-\frac{m}{4\pi}-\frac{T}{2\pi}\ln\left(1-e^{-m/T}\right)\,,
\end{equation}
cf. Ref.~\cite{Romatschke:2019ybu}\footnote{Note that the results for $I_{F,B}(m)$ can be obtained directly by writing $I_{B,F}(m)=\sumint_{\ P,\{P\}}\frac{1}{P^2+m^2}$ and performing the thermal sums, finding $I_{B,F}(m)=\int\frac{d^{2}p}{(2\pi)^2}\frac{1}{2\sqrt{p^2+m^2}}\left(1\pm 2 n_{\pm}(p)\right)$, with $n_{\pm}(x)=\frac{1}{e^{x/T}\mp 1}$ for bosons and fermions, respectively, cf. Ref.~\cite{Laine:2016hma}.}. It is tempting to interpret $\sigma^*$ as the in-medium fermion mass and $\sqrt{z^*}$ as the in-medium boson mass. Since the model does not break chiral symmetry, these in-medium masses vanish in the zero-temperature limit. Hence we may set
$z^*=m_B^2 T^2$, $\sigma^*=m_F T$ such that the above gap equations become
\begin{equation}
  \label{eq:gap}
  \frac{m_B^2}{4\lambda}=\frac{3 m_F^2}{4\lambda}+\frac{m_F^2}{2\pi}+\frac{m_F}{\pi}\ln\left(1+e^{-m_F}\right)\,,\quad
  \frac{m_F}{4\lambda}=-\frac{m_B}{4\pi}-\frac{1}{2\pi}\ln\left(1-e^{-m_B}\right)\,.
  \end{equation}
It is straightforward to verify that in the weak coupling limit $\lambda\rightarrow 0$, the solution to these equations is $m_F=m_B=0$, indicating vanishing thermal masses for both bosons and fermions.

At finite interaction strength, bosons and fermions develop non-vanishing thermal masses. At weak coupling $\lambda \ll 1$, these may be calculated in perturbation theory, but the gap equations (\ref{eq:gap}) can be solved numerically for any value of the interaction strength (see Fig.~\ref{fig1}).
Curiously, we find that in the strong coupling limit, the masses tend to
\begin{equation}
  \label{eq:strongmasses}
  \lim_{\lambda\rightarrow \infty} m_B=2\ln \frac{1+\sqrt{5}}{2}\simeq 0.96\,,\quad
  \lim_{\lambda\rightarrow \infty} m_F=0\,.
\end{equation}

\begin{figure*}[t]
  \centering
  \includegraphics[width=0.7\linewidth]{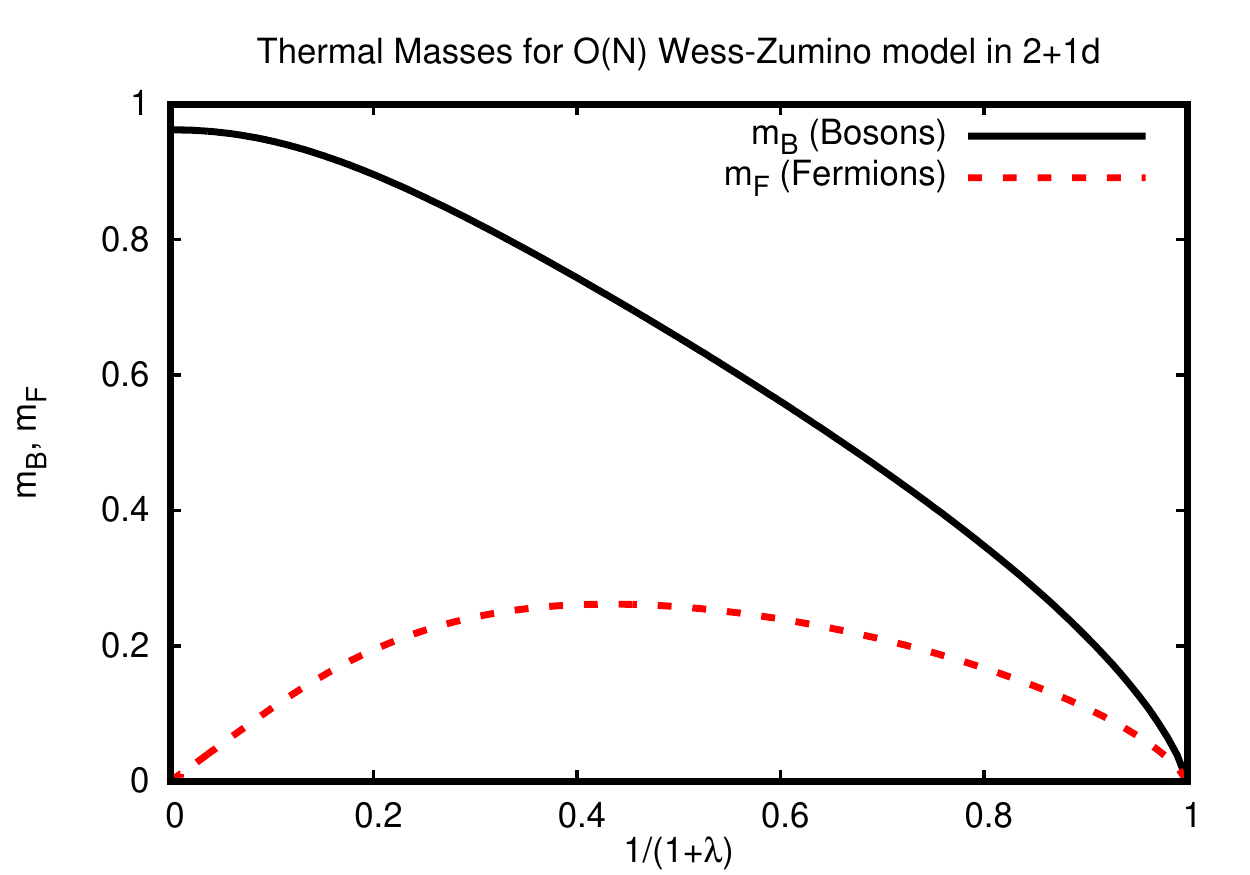}
  \caption{\label{fig1} Thermal masses (divided by temperature) in the O(N) Wess-Zumino model in 2+1d in the large N limit for all coupling values. Horizontal axis has been compactified in order to show the full range $\lambda \in [0,\infty)$. See text for details.}
\end{figure*}

\subsection{Thermodynamics}

The pressure $P$ (minus the free energy density) for the Wess-Zumino model in 2+1 dimensions can be calculated as
\begin{equation}
P=\frac{\ln Z}{\beta V}=N\left(-J_B(\sqrt{z^*})+J_F(\sigma^*)+\frac{z^* \sigma^*}{8\lambda}-\frac{\sigma^{* 3}}{8\lambda}\right)\,.
\end{equation}
The functions $J_B(m),J_F(m)$ may be evaluated by direct integration of the known forms of $I_B(m),I_F(m)$ in Eq.~(\ref{eq:is}). One finds
\begin{eqnarray}
  J_B(m)&=&-\frac{m^3}{6\pi}+\frac{m^2 T}{4\pi}\ln\frac{1-e^{m/T}}{1-e^{-m/T}}+\frac{m T^2}{2\pi}{\rm Li}_2\left(e^{m/T}\right)-\frac{T^3}{2\pi}{\rm Li}_3\left(e^{m/T}\right)\,,\nonumber\\
  J_F(m)&=&-\frac{m^3}{6\pi}+\frac{m^2 T}{4\pi}\ln\frac{1+e^{m/T}}{1+e^{-m/T}}+\frac{m T^2}{2\pi}{\rm Li}_2\left(-e^{m/T}\right)-\frac{T^3}{2\pi}{\rm Li}_3\left(-e^{m/T}\right)\,.
  \end{eqnarray}
Plugging in the expressions for the bosonic in-medium mass $\sqrt{z^*}=m_B T$ and fermionic in-medium mass $\sigma^*=m_F T$ allows evaluation of the pressure for all values of the interaction. However, the entropy density $s=\frac{\partial P}{\partial T}$ is a somewhat more convenient thermodynamic variable because the temperature derivatives of the in-medium masses $\sqrt{z^*},\sigma^*$ cancel as a result of the saddle point conditions (\ref{eq:saddles}). One finds
\begin{eqnarray}
  \label{eq:results}
  s&=&s_B(m_B)+s_F(m_F)\,,\\
  s_B(m_B)&=&\frac{N T^2}{4\pi}\left[m_B^3+m_B^2\ln\frac{1-e^{-m_B}}{\left(1-e^{m_B}\right)^3}-6 m_B {\rm Li}_2\left(e^{m_B}\right)+6 {\rm Li}_3\left(e^{m_B}\right)\right]\,,\nonumber\\
  s_F(m_F)&=&-\frac{N T^2}{4\pi}\left[m_F^3+m_F^2\ln\frac{1+e^{-m_F}}{\left(1+e^{m_F}\right)^3}-6m_F {\rm Li}_2\left(-e^{m_F}\right)+6 {\rm Li}_3\left(-e^{m_F}\right)\right]\,.\nonumber
\end{eqnarray}
In the limit where in-medium masses vanish, one has
\begin{equation}
  s_B(0)=\frac{3 N T^2}{2\pi} \zeta(3)\,,\quad
  s_F(0)=\frac{9 N T^2}{8\pi} \zeta(3)=\frac{3}{4}s_B(0)\,,
\end{equation}
where the factor $3/4$ is the known ratio of fermionic to bosonic degrees of freedom for a free theory in 2+1 dimensions. Thus, for a free theory,
\begin{equation}
  s_{\rm free}=s_B(0)+s_F(0)=\frac{7}{4}s_B(0)=\frac{21 N T^2}{8 \pi}\zeta(3)\,.
\end{equation}
For finite values of the interaction parameter, Eqns.~(\ref{eq:results}) may be evaluated numerically using the solutions for $m_B,m_F$ from the non-perturbative gap equations (\ref{eq:gap}) at any value of the coupling $\lambda$. The resulting dependence of the entropy density (scaled by $s_{\rm free}$) is shown in Fig.~\ref{fig2}.

In the strong coupling limit, $m_B\rightarrow 2\ln \frac{1+\sqrt{5}}{2}$ and the properties of the polylogarithms may be used to show \cite{Sachdev:1993pr,Drummond:1997cw}
\begin{equation}
  s_B\left(2\ln \frac{1+\sqrt{5}}{2}\right)=\frac{6 N T^2}{5 \pi} \zeta(3)=\frac{4}{5} s_B(0)\,.
\end{equation}
Since in the strong coupling limit the fermionic in-medium mass vanishes $m_F\rightarrow 0$, the entropy density in the strong coupling limit becomes
\begin{equation}
  \label{eq:strongentropy}
  \lim_{\lambda\rightarrow \infty} s=\frac{4}{5}s_B(0)+\frac{3}{4} s_B(0)=\frac{31}{35}\, s_{\rm free}\simeq 0.885714\, s_{\rm free}\,.
  \end{equation}

\begin{figure*}[t]
  \centering
  \includegraphics[width=0.7\linewidth]{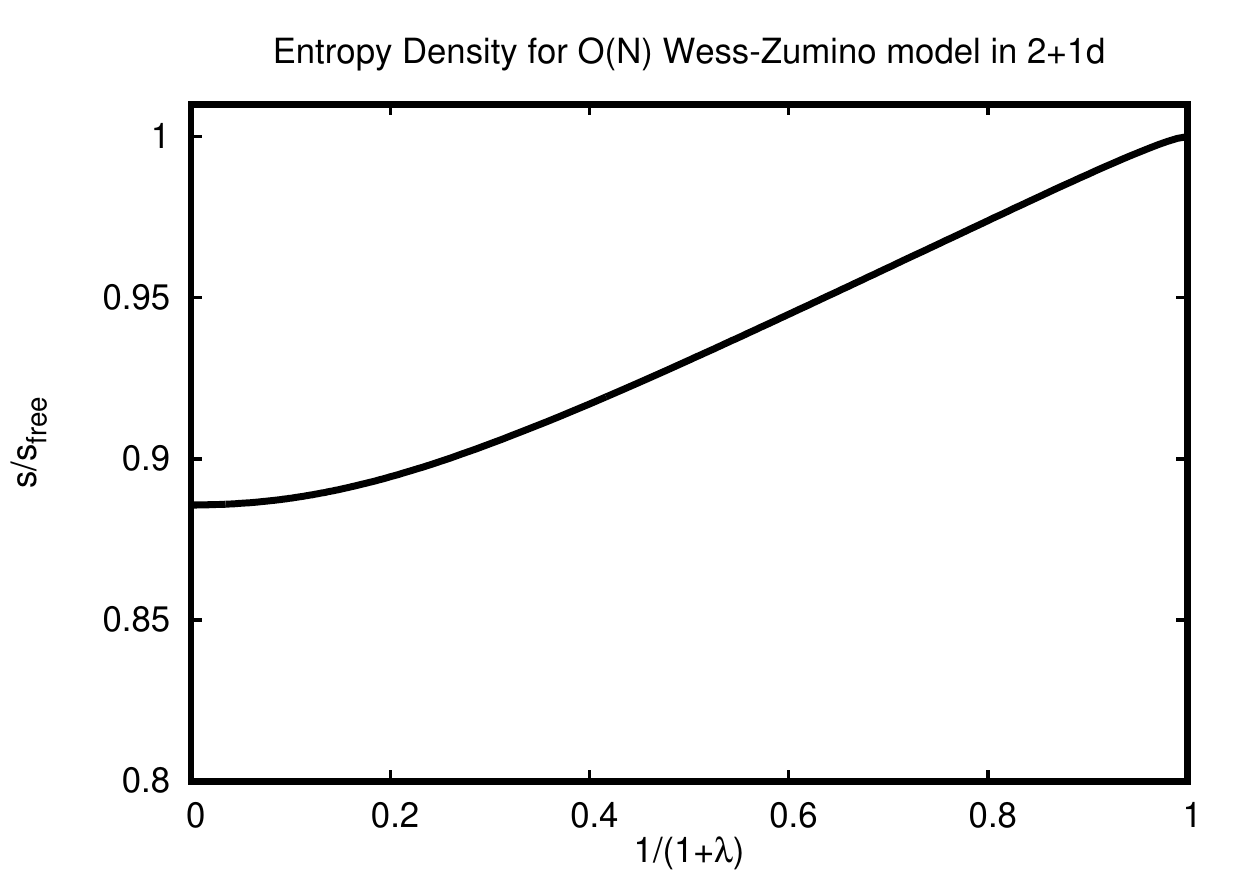}
  \caption{\label{fig2} Entropy density scaled by Stefan-Boltzmann limit in the O(N) Wess-Zumino model in 2+1d in the large N limit for all coupling values. Horizontal axis has been compactified in order to show the full range $\lambda \in [0,\infty)$. See text for details.}
\end{figure*}

\section{Strong Coupling Universality}

Taken by itself, the result that the large N Wess-Zumino model in 2+1 dimensions is a pure CFT with a strong-weak thermodynamic ratio of $\frac{31}{35}$ is perhaps not that relevant or enlightening. For this reason, it is interesting to study if the ratio of $\frac{31}{35}$ is valid more generally for a larger class of theories.

First, while the Wess-Zumino realizes ${\cal N}=1$ supersymmetry at zero temperature, note that the ratio $\frac{31}{35}$ does not depend on the theory to be supersymmetric (this is not expected given that finite temperature breaks supersymmetry anyway). To see this, simply change the coefficient of the bosonic interaction potential in (\ref{eq:seuclid}) from
\begin{equation}
\frac{16 \lambda^2}{2 N^2} \left(\phi_a\phi_a\right)^3\rightarrow \frac{\alpha}{2 N^2}\left(\phi_a\phi_a\right)^3\,,\quad \alpha>0\,,
\end{equation}
leaving the rest of the Lagrangian unchanged. Proceeding with the calculation as before, one finds that $m_F=0$ in the strong coupling limit. 
%
Therefore, while results for intermediate coupling values $\lambda$ are sensitive to the choice of $\alpha$, in the infinite coupling limit the results (\ref{eq:strongmasses}), (\ref{eq:strongentropy}) are valid also for $\alpha \neq 16\lambda^2$, and hence do not depend on the presence of supersymmetry.

In fact, the strong coupling results (\ref{eq:strongmasses}), (\ref{eq:strongentropy}) are valid for a much larger class of interaction potentials. Changing \begin{equation}
\frac{16\lambda^2}{2 N^2} \left(\phi_a\phi_a\right)^3\rightarrow N\times U\left(\frac{\phi_a \phi_a}{N}\right)\,,\quad 
\end{equation}
we can consider arbitrary interaction potentials $U$ including the possibility of adding relevant operators. The only conditions on $U(x)$ are that there is no spontaneous mass generation which would break the CFT, and that its argument is non-negative (since $x=\frac{\phi_a\phi_a}{N}$). The former condition requires that the potential only has one trivial minimum located at $x=0$ or $U^\prime(0)=0$, $U^{\prime\prime}(0)>0$.

For any such potential $U$, the relevant gap equation gets modified to
\begin{equation}
   \frac{m_B^2}{4\lambda}=\frac{1}{2\lambda T^2}U^\prime\left(\frac{m_F T}{4 \lambda}\right)+\frac{m_F^2}{2\pi}+\frac{m_F}{\pi}\ln\left(1+e^{-m_F}\right)\,.
\end{equation}
For $\lambda\rightarrow \infty$, $m_B\rightarrow 2 \ln \frac{1+\sqrt{5}}{2}$ as before, such that the lhs of this equation tends to zero. Since $U^{\prime\prime}(0)>0$ and the only minimum of the potential is located at $x=0$, this implies that the rhs of the above equation is positive definite for all $m_F>0$, leading to $m_F=0$ as the only possible solution.

As a consequence, the strong coupling results (\ref{eq:strongmasses}), (\ref{eq:strongentropy}) are universally true for a large class of interaction potentials $U(x)$ in the large N limit. 

\subsection{Changing the balance between fermions and bosons}

As is easy to see, the above argument trivially extends to cases where the field content of the theory is modified to contain multiple copies of the scalars and fermions as long as they come in equal number. This establishes the universality of the strong-weak thermodynamic ratio of $\frac{31}{35}$ for a large class of 2+1 dimensional pure CFTs with equal number of bosonic and fermionic degrees of freedom in the large N limit. 

But what about changing the balance between fermionic and bosonic degrees of freedom? For this reason, let us consider $B$ N-component scalars coupled to $F$ N-component Majorana spinors in 2+1 dimensions with a Euclidean action given by
\begin{equation}
  \label{eq:seuclidmod}
  S_E=\int d^3x \left(\frac{B}{2}\partial_\mu \phi_a \partial_\mu \phi_a
  +\frac{F}{2}\bar\psi_a \slashed{\partial}\psi_a
  +B\times N\times U\left(\frac{\phi_a\phi_a}{N}\right) + \frac{2\lambda}{N} {\rm min}(F,B)\times \bar\psi_a \psi_a \phi_b\phi_b\right)\,.
\end{equation}
Proceeding as above leads to a partition function for this theory which is given by
\begin{equation}
  Z=\int d\sigma d\zeta e^{-\beta V N\left(
B U\left(\frac{\sigma}{2\lambda}\right)-\frac{i B \zeta \sigma}{4\lambda}+B J_B\left(\sqrt{i \zeta}\right)-{\rm min}(F,B)J_F(\sigma)-(F-{\rm min}(F,B))J_F(0)
    \right)}\,,
  \end{equation}
and the gap equations become
\begin{eqnarray}
  \frac{m_B^2}{4\lambda}&=&\frac{U^\prime\left(\frac{m_F T}{4\lambda}\right)}{2 \lambda T^2}+\frac{{\rm min}(F,B)}{B}\left(\frac{m_F^2}{2\pi}+\frac{m_F}{\pi}\ln\left(1+e^{-m_F}\right)\right)\,,\nonumber\\
  \frac{m_F}{4\lambda}&=&-\frac{m_B}{4\pi}-\frac{1}{2\pi}\ln\left(1-e^{-m_B}\right)\,.
  \end{eqnarray}
Repeating the argument from above, this implies that all $B$ copies of the N-component bosonic degrees of freedom develop a mass $m_B=2 \ln\frac{1+\sqrt{5}}{2}$ in the limit $\lambda\rightarrow \infty$, whereas all $F$ copies of the N-component fermions have $m_F=0$ in that limit. (Note that some fermions, namely ${\rm min(F,B)}$ of them, develop a non-zero mass for intermediate coupling, while the others are always free.)

Therefore, the strong-weak thermodynamic ratio for this class of theories with B N-component scalars and F N-component fermions is given by
\begin{equation}
\lim_{\lambda\rightarrow \infty}\frac{s}{s_{\rm free}}=\frac{\frac{4}{5}+\frac{3 F}{4 B}}{1+\frac{3 F}{4 B}}\,,
\end{equation}
which is a function that monotonically increases with the ratio of fermionic to bosonic degrees of freedom $F/B$. For $F=B$, one recovers $\frac{31}{35}$, as above. Interesting limits of the ratio $\frac{s}{s_{\rm free}}$ are no fermions (or, equivalently, an infinite number of bosons with only a finite number of fermions), for which $\frac{s}{s_{\rm free}}=\frac{4}{5}$, whereas for the opposite limit of no bosons (or an infinite number of fermions), $\frac{s}{s_{\rm free}}=1$.

Thus we find that for a large class of theories with different field content and different interactions the thermodynamic ratio is bounded by
\begin{equation}
  \frac{4}{5}\leq \lim_{\lambda\rightarrow \infty}\frac{s}{s_{\rm free}}\leq 1\,,
\end{equation}
in the large N limit in 2+1 dimensions.

\section{Summary and Conclusions}

In this work, we have considered the thermodynamics of pure CFTs in 2+1 dimensions with interacting fermions and bosons in the large N limit. We solved the supersymmetric O(N) Wess-Zumino model analytically at finite temperature for all coupling values in the large N limit. Our results show that the entropy density monotonically decreases from its Stefan-Boltzmann value $s_{\rm free}$ at weak coupling to $\frac{31}{35}\times s_{\rm free}$ at infinite coupling.

At intermediate values of the coupling, the thermodynamic ratio $\frac{s}{s_{\rm free}}$ depends on the details of the interaction. However, we found that the infinitely strong to weak coupling thermodynamic ratio of $\frac{31}{35}$ is universal for a large class of interaction potentials with equal number of fermionic and bosonic degrees of freedom in 2+1 dimensions, with and without supersymmetry. Furthermore, we found that for unequal numbers of fermionic and bosonic degrees of freedom the corresponding weak-strong thermodynamic ratio is bounded by the purely bosonic result 4/5 from Ref.~\cite{Romatschke:2019ybu} from below, and unity above.

Many open questions remain, such as finite N corrections to the ratios 4/5 and 31/35 within the class of theories considered here, or if pure CFTs with gauge fields exist that can be solved for all interaction strengths. Extensions of this work to higher odd dimensions is feasible and could be considered along the lines of Ref.~\cite{Filothodoros:2018pdj}. Examples of solvable field theories with simple known gravity duals are rare, cf. Refs.~\cite{Hanada:2013rga,Berkowitz:2016jlq}. Therefore, another direction to consider could be studies of possible gravitational duals to the CFTs considered here, such as the gravity dual to the O(N) Wess-Zumino model in 2+1 dimensions along the lines of Ref.~\cite{Klebanov:2002ja}. We leave these questions to future work.

  \section{Acknowledgments}

  This work was supported by the Department of Energy, DOE awards No DE-SC0017905 (PR) and DE-SC0010005 (OD). We would like to thank S.P. de Alwis and T. DeGrand for helpful discussions.

\bibliographystyle{JHEP}
\bibliography{cft}

\providecommand{\href}[2]{#2}\begingroup\raggedright\begin{thebibliography}{10}

\bibitem{Maldacena:1997re}
J.~M. Maldacena, {\it {The Large N limit of superconformal field theories and
  supergravity}},  {\em Int. J. Theor. Phys.} {\bf 38} (1999) 1113--1133,
  [\href{http://arxiv.org/abs/hep-th/9711200}{{\tt hep-th/9711200}}]. [Adv.
  Theor. Math. Phys.2,231(1998)].

\bibitem{Gubser:1998nz}
S.~S. Gubser, I.~R. Klebanov, and A.~A. Tseytlin, {\it {Coupling constant
  dependence in the thermodynamics of N=4 supersymmetric Yang-Mills theory}},
  {\em Nucl. Phys.} {\bf B534} (1998) 202--222,
  [\href{http://arxiv.org/abs/hep-th/9805156}{{\tt hep-th/9805156}}].

\bibitem{Belavin:1984vu}
A.~A. Belavin, A.~M. Polyakov, and A.~B. Zamolodchikov, {\it {Infinite
  Conformal Symmetry in Two-Dimensional Quantum Field Theory}},  {\em Nucl.
  Phys.} {\bf B241} (1984) 333--380.

\bibitem{inprep}
P.~Romatschke, {\it {An example of pure CFT thermodynamics for all couplings
  and all N in 1+1 dimensions}},  \href{http://arxiv.org/abs/in
  preparation}{{\tt in preparation}}.

\bibitem{Drummond:1997cw}
I.~T. Drummond, R.~R. Horgan, P.~V. Landshoff, and A.~Rebhan, {\it {Foam
  diagram summation at finite temperature}},  {\em Nucl. Phys.} {\bf B524}
  (1998) 579--600, [\href{http://arxiv.org/abs/hep-ph/9708426}{{\tt
  hep-ph/9708426}}].

\bibitem{Sachdev:1993pr}
S.~Sachdev, {\it {Polylogarithm identities in a conformal field theory in
  three-dimensions}},  {\em Phys. Lett.} {\bf B309} (1993) 285--288,
  [\href{http://arxiv.org/abs/hep-th/9305131}{{\tt hep-th/9305131}}].

\bibitem{Romatschke:2019ybu}
P.~Romatschke, {\it {Finite temperature CFT results for all couplings: O(N)
  model in 2+1 dimensions}},  \href{http://arxiv.org/abs/1904.09995}{{\tt
  arXiv:1904.09995}}.

\bibitem{Wess:1974tw}
J.~Wess and B.~Zumino, {\it {Supergauge Transformations in Four-Dimensions}},
  {\em Nucl. Phys.} {\bf B70} (1974) 39--50.

\bibitem{Laine:2016hma}
M.~Laine and A.~Vuorinen, {\it {Basics of Thermal Field Theory}},  {\em Lect.
  Notes Phys.} {\bf 925} (2016) pp.1--281,
  [\href{http://arxiv.org/abs/1701.01554}{{\tt arXiv:1701.01554}}].

\bibitem{Filothodoros:2018pdj}
E.~G. Filothodoros, A.~C. Petkou, and N.~D. Vlachos, {\it {The fermion-boson
  map for large $d$}},  {\em Nucl. Phys.} {\bf B941} (2019) 195--224,
  [\href{http://arxiv.org/abs/1803.05950}{{\tt arXiv:1803.05950}}].

\bibitem{Hanada:2013rga}
M.~Hanada, Y.~Hyakutake, G.~Ishiki, and J.~Nishimura, {\it {Holographic
  description of quantum black hole on a computer}},  {\em Science} {\bf 344}
  (2014) 882--885, [\href{http://arxiv.org/abs/1311.5607}{{\tt
  arXiv:1311.5607}}].

\bibitem{Berkowitz:2016jlq}
E.~Berkowitz, E.~Rinaldi, M.~Hanada, G.~Ishiki, S.~Shimasaki, and P.~Vranas,
  {\it {Precision lattice test of the gauge/gravity duality at large-$N$}},
  {\em Phys. Rev.} {\bf D94} (2016), no.~9 094501,
  [\href{http://arxiv.org/abs/1606.04951}{{\tt arXiv:1606.04951}}].

\bibitem{Klebanov:2002ja}
I.~R. Klebanov and A.~M. Polyakov, {\it {AdS dual of the critical O(N) vector
  model}},  {\em Phys. Lett.} {\bf B550} (2002) 213--219,
  [\href{http://arxiv.org/abs/hep-th/0210114}{{\tt hep-th/0210114}}].

\end{thebibliography}\endgroup
\end{document}